\documentclass{article}

\usepackage{microtype}
\usepackage{graphicx}
\usepackage{subfigure}
\usepackage{booktabs} 
\usepackage{multicol} 
\usepackage{adjustbox} 
\usepackage{float} 
\usepackage{xcolor} 
\usepackage{parskip}

\usepackage{hyperref}

\usepackage[accepted]{icml2023} 

\icmltitlerunning{What can we learn from Data Leakage and Unlearning for Law?}

\begin{document}

\twocolumn[
\icmltitle{What can we learn from Data Leakage and Unlearning for Law?}





\begin{icmlauthorlist}
\icmlauthor{Jaydeep Borkar}{yyy}
\end{icmlauthorlist}

\icmlaffiliation{yyy}{Khoury College of Computer Sciences, Northeastern University, Boston MA, USA}

\icmlcorrespondingauthor{Jaydeep Borkar}{borkar.j@northeastern.edu}

\icmlkeywords{Machine Learning, ICML}

\vskip 0.3in
]



\printAffiliationsAndNotice{\icmlEqualContribution} 

\section{Introduction} 
Large Language Models (LLMs) have a privacy concern because they memorize training data and leak it during text generation which is often described as data leakage. The memorized data might include personally identifiable information (PII) like emails and phone numbers as well as some copyrighted content. There has been important work on studying memorization and data leakage for the general purpose pre-trained or foundation models \cite{carlini2021, carlini2023quantifying, lehman-etal-2021-bert, pi, deduplication, kandpal2022deduplicating, biderman2023emergent}. However, in a real-world scenario, a small organization or company that doesn't have enough computational resources to train an LLM on its own data will prefer fine-tuning a pre-trained model on its domain-customized dataset as it is computationally much cheaper. Fine-tuning refers to adapting a pre-trained model for a specific domain and tasks using some additional data\footnote{https://genlaw.github.io/glossary.html}. So far, little attention has been given to understanding memorization and data leakage for fine-tuned models. This makes exploring memorization in fine-tuned models important as the fine-tuning datasets might potentially include PII and other information that could be leaked. 

Pre-trained models can be obtained from organizations that offer LLM-as-a-service for fine-tuning. The dataset used during pre-training could be private or proprietary data that may not be intended to be publicly available. Hence, it's also important to ensure that the pre-training data is not leaked through the fine-tuned models after fine-tuning. There has been some work on understanding memorization for fine-tuned models where the authors insert multiple copies of a secret sequence in the dataset and then evaluate memorization for those sequences \cite{mireshghallah-etal-2022-empirical}. In this work, we show that a fine-tuned model can potentially leak fine-tuning as well as the pre-training data through text generation. 

\begin{figure}[htp]
  \centering
  \includegraphics[width=1\columnwidth]{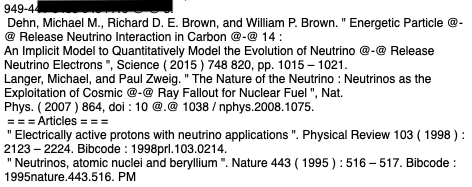}
  \caption{A potential phone number (that doesn't belong to Wiki-Text103) being leaked after querying GPT-2 fine-tuned on Wiki-Text103. This indicates that examples memorized during pre-training could be leaked even by the fine-tuned model.}
  \label{fig:mobile}
  \vspace{-10pt}
\end{figure}

As we discussed earlier, the dataset used for fine-tuning might contain private information like PII. The company that fine-tunes the model might implement various solutions to prevent the leakage of PII. One such solution is unlearning where specific data points are explicitly removed from the dataset and the model is re-trained or fine-tuned again on the new dataset \cite{unlearning, bourtoule2020machine}. The company can perform unlearning either to remove the data points that are highly vulnerable to leakage or in order to comply with the ``\textit{right to be forgotten}" policy where the users can request their data to be removed from the dataset \cite{right-forgotten}. We find that once we unlearn the data points that are highly vulnerable to leakage, a new set of data points that were previously safe become vulnerable to leakage.

The property of previously safe data points becoming vulnerable to leakage after unlearning and leakage of pre-training and fine-tuning data through fine-tuned models can pose significant privacy and legal concerns for companies that use LLMs to offer services. We hope that these preliminary results will start an interdisciplinary discussion within Artificial Intelligence and law communities regarding the need for policies to tackle these issues. According to the best of our knowledge, this is the first work to study the leakage of pre-training and fine-tuning data in fine-tuned models through text generation, the impact of unlearning in large language models on the privacy of data points, and the overall connection to law and policy. 

\newpage
In a nutshell, our contributions can be summarized by the following takeaways: 
\begin{enumerate} 
   \setlength\itemsep{0em}
    \item Fine-tuned models can leak data from the fine-tuning dataset including PII such as email addresses. 
    \item Unlearning data points of specific users who are vulnerable to data extraction could potentially jeopardize the privacy of remaining data points in the dataset. 
    \item If an organization trains its own LLM from scratch on proprietary training data and makes it available to others \textit{only} for fine-tuning, the fine-tuned models could potentially leak the proprietary training data. 
    
\end{enumerate}

\section{Related Work} 
Carlini et al. were the first to show that generative text models suffer from unintended memorization which can have privacy concerns \cite{secret-sharer}. It has been found that large language models like GPT-2 memorize and leak training data \cite{carlini2021}. The amount of memorized data can also be quantified \cite{carlini2023quantifying}. There have been works that attempt to extract training data from BERT \cite{bert} trained on clinical notes \cite{lehman-etal-2021-bert} and studying memorization of PII \cite{pi, lukas2023analyzing}. NLP fine-tuning methods have been found to show a memorization behavior \cite{mireshghallah-etal-2022-empirical}. Mireshghallah et al. design a membership inference attack to predict the membership of points for Masked Languaged Models (MLMs) \cite{mireshghallah2022quantifying}. Carlini et al. talk about the impact of unlearning on the privacy of remaining points in image datasets \cite{onion}. Unlearning refers to the removal of specific data points from the training dataset \cite{unlearning, bourtoule2020machine}. Various legislations such as the General Data Protection Legislation (GDPR) in European Union \cite{eu}, the California Consumer Privacy Act in the United States \cite{cali},  and PIPEDA privacy legislation in Canada \cite{canada} talk about the \textit{right to be forgotten} \cite{right-forgotten} policy where the users have the right for their data to be deleted from the models.

\section{Experiments and Preliminary Results}
\label{section:setup}
Our experimental set-up is a subset replica of what Carlini et al. have proposed to ensure that we study memorization under a similar setting \cite{carlini2021}\footnote{\url{https://github.com/ftramer/LM\_Memorization}}. We generate 2000 samples (256 tokens each) in total using the top-k sampling method (k=40) \cite{top-k} by prompting the model in the following ways: (1) Prompting the model with the start-of-the-sequence token (2) Prompting the model with random ten tokens from the Common Crawl\footnote{\url{https://commoncrawl.org/}} for each sample. Further, we sort the generated samples by using metrics like perplexity and zlib entropy \cite{zlib}\footnote{zlib entropy can be calculated as the length of the compressed data (bytes)}. To evaluate memorization for the fine-tuning dataset, we perform a search to find common n-grams between generated samples and the dataset. To check for data memorized during the pre-training phase, we simply perform an internet search for that sample as GPT-2 is trained on data scraped from the internet. Section \ref{section:1} talks about fine-tuning data leakage, Section \ref{section:2} shows that fine-tuned models can leak pre-training data, and Section \ref{section:3} demonstrates how mitigation methods such as unlearning can have an adverse effect on the overall privacy. 

\subsection{Extracting fine-tuning data from fine-tuned model}
\label{section:1}
We generate samples from GPT-2 large fine-tuned on WikiText-103 from Hugging Face \cite{alon2022neuro, wikitext} using the methods discussed in Section \ref{section:setup}. We were able to extract short sequences that included named entities such as a list (ordered in a particular way) of musicians, celebrities, organizations, museums, songs, universities, URLs, etc. For longer sequences, we were able to extract sequences where 100+ tokens were memorized (see Table \ref{tab:tab1}). Even though the WikiText-103 dataset is publicly available and doesn't contain any sensitive information as such, we can learn something from the results about the type of memorization one can expect if the dataset has private and copyrighted content. In Section \ref{section:3}, we show that if the fine-tuning dataset has sensitive data like PII in it then the fine-tuned model can potentially leak it. 

\begin{table}[htbp!]
  \centering
  \vspace{0pt}
  \small 
  \scalebox{1}{%
    \begin{tabular}{|p{0.8\linewidth}|}
      \hline
      \textbf{The Boat Race is a side @-@ by @-@ side rowing competition between the University of Oxford ( sometimes referred to as the " Dark Blues " ) and the University of Cambridge ( sometimes referred to as the " Light Blues " ). The race was first held in 1829, and since 1845 has taken place on the 4 @.@ 2 @-@ mile ( 6 @.@ 8 km ) Championship Course on the River Thames in southwest London. The rivalry is a major point of honour between the two universities} ; it is followed throughout the United Kingdom  \\
      \hline
    \end{tabular}%
    
   }
   \caption{Memorized sample from GPT-2 fine-tuned on Wikitext103. All the text in bold is memorized.}
    \label{tab:tab1}
\end{table}

\subsection{Extracting pre-training data from fine-tuned model}
\label{section:2}
We observe that not only do fine-tuned models leak data from their fine-tuning dataset, but they also leak data that was memorized during the pre-training phase. We generate samples\footnote{since we use tokens from common crawl for prompts we assume that the attacker has access to a dataset with similar distribution to craft prompts.} from the fine-tuned model using methods discussed in Section \ref{section:setup} and sort them according to the perplexity of pre-trained GPT-2. We were able to extract content like actual phone number (see Figure \ref{fig:mobile}), URLs, Twitter handle, 13-digit alpha-numeric tracking numbers, 8-digit PMID number of articles on PubMed, an 8-digit company ID that results in information about the company's employees on the UK government's website, numbers for latitude and longitude which resulted in actual location after performing reverse geocoding, etc (see section \ref{section:pretrainleak} for some of these examples). None of these extracted examples were present in the Wikitext103 dataset which we used for fine-tuning but we could find them through a simple internet search. This implies that they were memorized during the pre-training phase and then later inherited by the fine-tuned model. Leakage of pre-training data can also be linked to model attribution where one could trace down the base model based on the output of the fine-tuned model \cite{merkhofer2023mlmac}. 


Our results indicate that both the pre-training and fine-tuning data could be leaked simultaneously by the fine-tuned model during text generation. Hence, it becomes necessary to identify from which dataset the memorized data (including PII) is coming from in order to apply mitigation strategies. On analyzing the structure of our memorized samples, we observed that the first few lines contained the text that belonged to the pre-training data and that is where we found the pre-training memorized data.

\subsection{Unlearning the extracted points and its impact on the overall privacy}
\label{section:3}
Companies can delete data points that are at a higher risk of extraction or in order to comply with the \textit{right to be forgotten} policy \cite{bourtoule2020machine, unlearning, right-forgotten}. Carlini et al. were the first to show that once the most vulnerable points are unlearned in image datasets like CIFAR-10, a new set of previously safe neighboring points get memorized \cite{onion}. We study this phenomenon for PII present in text datasets of large language models. We embed email addresses from the Enron dataset\footnote{\url{https://www.cs.cmu.edu/~enron/}} in the WikiText-2 dataset \cite{wikitext} and fine-tune GPT-2 small on it \cite{model, data}. We generate samples using the method described in section \ref{section:setup}. Initially, the dataset had 6523 email addresses and we were able to extract 44 out of them. We unlearn these 44 email addresses by removing them from the dataset\footnote{We perform exact unlearning where we remove the data points explicitly from the dataset} and fine-tuned GPT-2 on the unlearned dataset. After unlearning, we found that 20 new email addresses got leaked by the model which were previously safe. We call them \textit{onion points} taking inspiration from previous works on image datasets where they call it an onion effect \cite{onion}. 

\begin{figure}[htp]
\vspace{-12pt}
  \subfigure[zlib entropy and perplexity of GPT-2]{\includegraphics[scale=0.26]{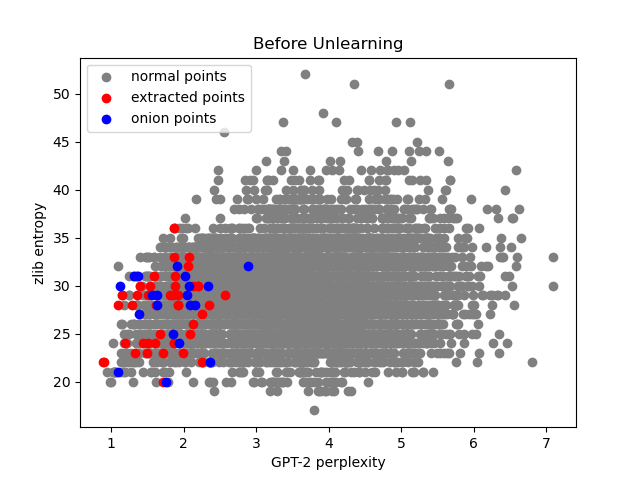}\label{fig:fig1}}
  \subfigure[Embeddings of email addresses]{\includegraphics[scale=0.26]{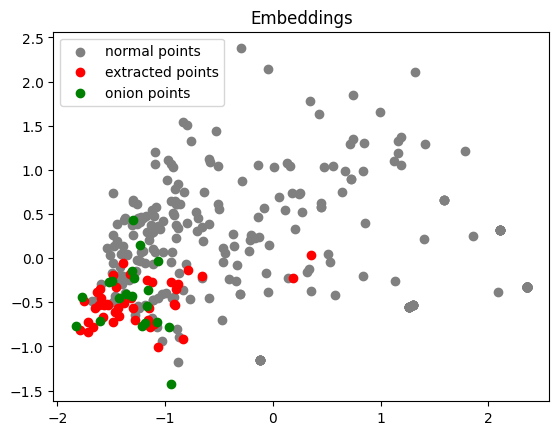}\label{fig:fig2}}
  \vspace{-12pt}
\end{figure}

In Figure \ref{fig:fig1}, we can see that the initially extracted 44 email addresses (red) and the 20 onion points (blue) are very close to each other and have a lower perplexity. The perplexity of onion points decreases after unlearning, which indicates that they were memorized. In Figure \ref{fig:fig2}, the embeddings\footnote{We use gensim's glove-wiki-gigaword-50 model to find embeddings \cite{glove, gensim}} for the initially extracted 44 email addresses (red) and 20 onion points (green) are very close to each other indicating that they have some similarities. We can say that the points that will be at a higher risk of getting vulnerable to leakage after unlearning will be usually the neighboring points. It's worthwhile to study this behavior for larger datasets and for different types of PII. Section \ref{section:unlearning} shows an example that is leaked during text generation. 

\section{Conclusion}
The leakage of pre-training and fine-tuning data (and PII) through fine-tuned models and the consequences of unlearning on overall privacy can potentially cause legal and privacy concerns for companies and organizations that provide LLMs as-as-service. We believe that our findings will provide insights to folks from Artificial Intelligence and Law communities into the need for necessary measures like dynamic privacy auditing and checking for memorization of proprietary training data and personal information in LLMs as they get deployed in the real world. 

\section{Acknowledgements}
We would like to thank Fatemehsadat Mireshghallah, Pin-Yu Chen, and anonymous reviewers for their feedback and insightful discussions on this work. 

\bibliography{main}
\bibliographystyle{icml2021}

\section{Appendix}
\subsection{Examples present in pre-training dataset and leaked by the fine-tuned model}
\label{section:pretrainleak}
These are some examples that are present in the training dataset of the original GPT-2 model\footnote{We do an internet search for the memorized examples as GPT-2 is originally trained on data scraped from the internet.} but get leaked even after it is fine-tuned on the Wiki-Text103 dataset. Thus, the leakage of pre-training data can occur even through fine-tuned models in addition to the original model. All of these examples are not present in the Wiki-Text103 dataset but very likely to be present in GPT-2's original training dataset. If we analyze the structure of these examples, we can see that the first half seems to be coming from the pre-training dataset and the latter half from the fine-tuning dataset (usually starting with ``=" patterns which indicates headings in the Wiki-Text103 dataset).

\begin{figure}[ht!]
  \centering
  \includegraphics[width=1\columnwidth]{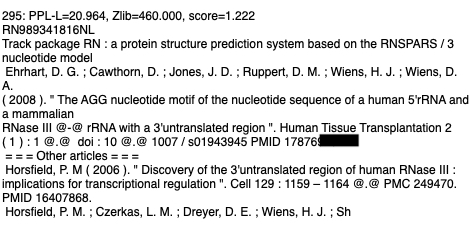}
  \caption{PMID number of an article on PubMed.}
  \label{fig:myimage}
\end{figure}

\begin{figure}[ht!]
  \centering
  \includegraphics[width=1\columnwidth]{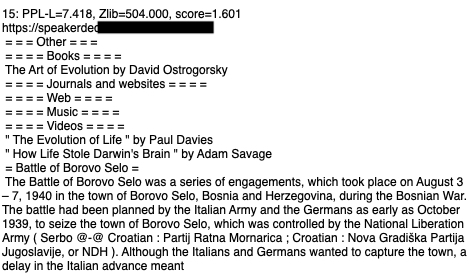}
  \caption{URL}
  \label{fig:myimage}
\end{figure}

\begin{figure}[ht!]
  \centering
  \includegraphics[width=1\columnwidth]{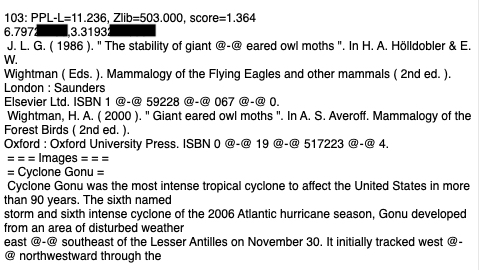}
  \caption{Numbers that potentially contain coordinates for the longitude and latitude of a place in Nigeria after removing the last two and last four digits from each.}
  \label{fig:myimage}
\end{figure}

\begin{figure}[ht!]
  \centering
  \includegraphics[width=1\columnwidth]{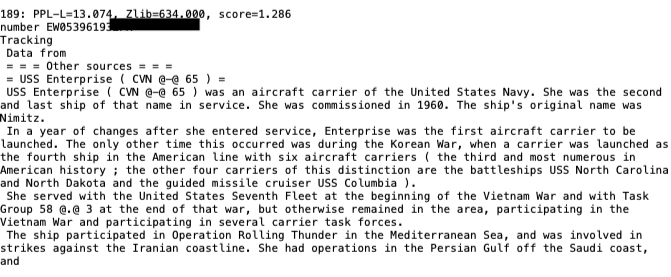}
  \caption{Some tracking number}
  \label{fig:myimage}
\end{figure}

\begin{figure}[ht!]
  \centering
  \includegraphics[width=1\columnwidth]{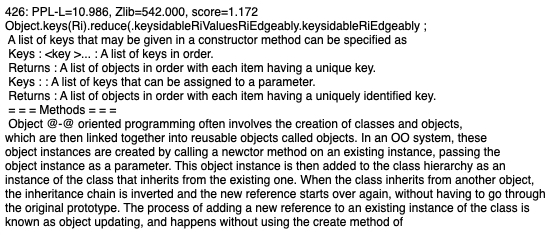}
  \caption{Snippet of code.}
  \label{fig:myimage}
\end{figure}



\end{document}